\documentclass[aps, prb, reprint, superscriptaddress, showpacs]{revtex4-2}
\usepackage{amssymb}
\usepackage{amsmath}
\usepackage{bm}
\usepackage{graphicx}
\usepackage{color}
\usepackage[dvipdfmx]{hyperref}
\hypersetup{
    colorlinks=true,
    linkcolor=blue,
    filecolor=magenta,      
    urlcolor=cyan,
    pdftitle={manuscript},
    pdfpagemode=FullScreen,
    }

\allowdisplaybreaks

\newcommand{\lBEDSe}{$\lambda$-BEDSe}
\newcommand{\lBETS}{$\lambda$-BETS}
\newcommand{\lBEDSeFe}{$\lambda$-(BEDSe-TTF)$_2$FeCl$_4$}
\newcommand{\lFe}{$\lambda$-(BETS)$_2$FeCl$_4$}

\begin{document}

\title{Dynamics of ethylene groups and hyperfine interactions between donor and anion molecules in $\lambda$-type organic conductors studied by $^{69,71}$Ga-NMR spectroscopy}

\author{N.~Yasumura}
\affiliation{Graduate School of Science and Engineering, Saitama University, Saitama, 338-8570, Japan}
\author{T.~Kobayashi}
\altaffiliation{tkobayashi@phy.saitama-u.ac.jp}
\affiliation{Graduate School of Science and Engineering, Saitama University, Saitama, 338-8570, Japan}
\affiliation{Research and Development Bureau, Saitama University, Saitama 338-8570, Japan}
\author{H.~Taniguchi}
\affiliation{Graduate School of Science and Engineering, Saitama University, Saitama, 338-8570, Japan}
\author{S.~Fukuoka}
\author{A.~Kawamoto}
\affiliation{Department of Condensed Matter Physics, Graduate School of Science, Hokkaido University, Sapporo 060-0810, Japan}

\date{\today}

\begin{abstract}
We present the results of $^{69,71}$Ga-NMR measurements on an organic antiferromagnet $\lambda$-(BEDSe-TTF)$_2$GaCl$_4$ [BEDSe-TTF=bis(ethylenediseleno)tetrathiafulvalene], with comparison to reports on $\lambda$-(BETS)$_2$GaCl$_4$ [BETS=bis(ethylenedithio)tetraselenafulvalene] [T. Kobayashi {\it et al}., \href{https://doi.org/10.1103/PhysRevB.102.235131}{Phys. Rev. B {\bf 102}, 235131 (2020)}]. 
We found that the dynamics of two crystallographically independent ethylene groups induce two types of quadrupolar relaxation in the high-temperature region. 
As the ethylene motion freezes, hyperfine (HF) interactions develop between $\pi$ spin and Ga nuclear spin below $100$~K, and thereby magnetic fluctuations of the $\pi$-spin system are detected even from the Ga site. 
The HF interaction in $\lambda$-(BETS)$_2$GaCl$_4$ was more than twice as large as in $\lambda$-(BEDSe-TTF)$_2$GaCl$_4$, implying that the short contacts between Cl atoms and the chalcogens of fulvalene part are essential for the transferred HF interaction. 
We propose that NMR using nuclei in anion layers is useful for studying interlayer interactions in organic conductors, which have not been studied experimentally.
In addition, because the mechanism of the transferred HF interaction is considered to be the same as $\pi$--$d$ interaction in isostructural Fe-containing $\lambda$-type salts, our findings aid in the understanding of their physical properties. 
\end{abstract}

\maketitle

\section{Introduction}
Most ET-based organic conductors are regarded as quasi-two-dimensional (Q2D) electronic systems, as well as cuprate and iron-based superconductors, where ET denotes bis(ethylenedithio)tetrathiafulvalene. 
They  often exhibit unconventional superconductivity in the vicinity of  magnetically ordered phases \cite{Bennemann2008}. 
To further understand these Q2D electronic systems, the role of the interlayer interaction has been discussed since long-range ordered states, such as superconductivity and magnetic ordering, require three-dimensional interaction. 
Electron spin resonance measurements in representative $\kappa$-type ET salts suggest that the Fermi liquid and Mott insulating states of these materials are formed below the temperature at which interlayer electron hopping and exchange interaction develop, respectively \cite{Antal2011, Antal2012a}. 
The magnetic structures of two $\kappa$-type salts have recently been revealed. 
Although they had been considered to be located in the same antiferromagnetic (AF) phase, their magnetic structures were found to be different each other \cite{Ishikawa2018, Oinuma2020}. 
It is concluded that the difference in magnetic structure is due to difference in the sign of the interlayer magnetic interaction. 

\begin{figure}[tbp]
\begin{center}
\includegraphics[width=8.5cm]{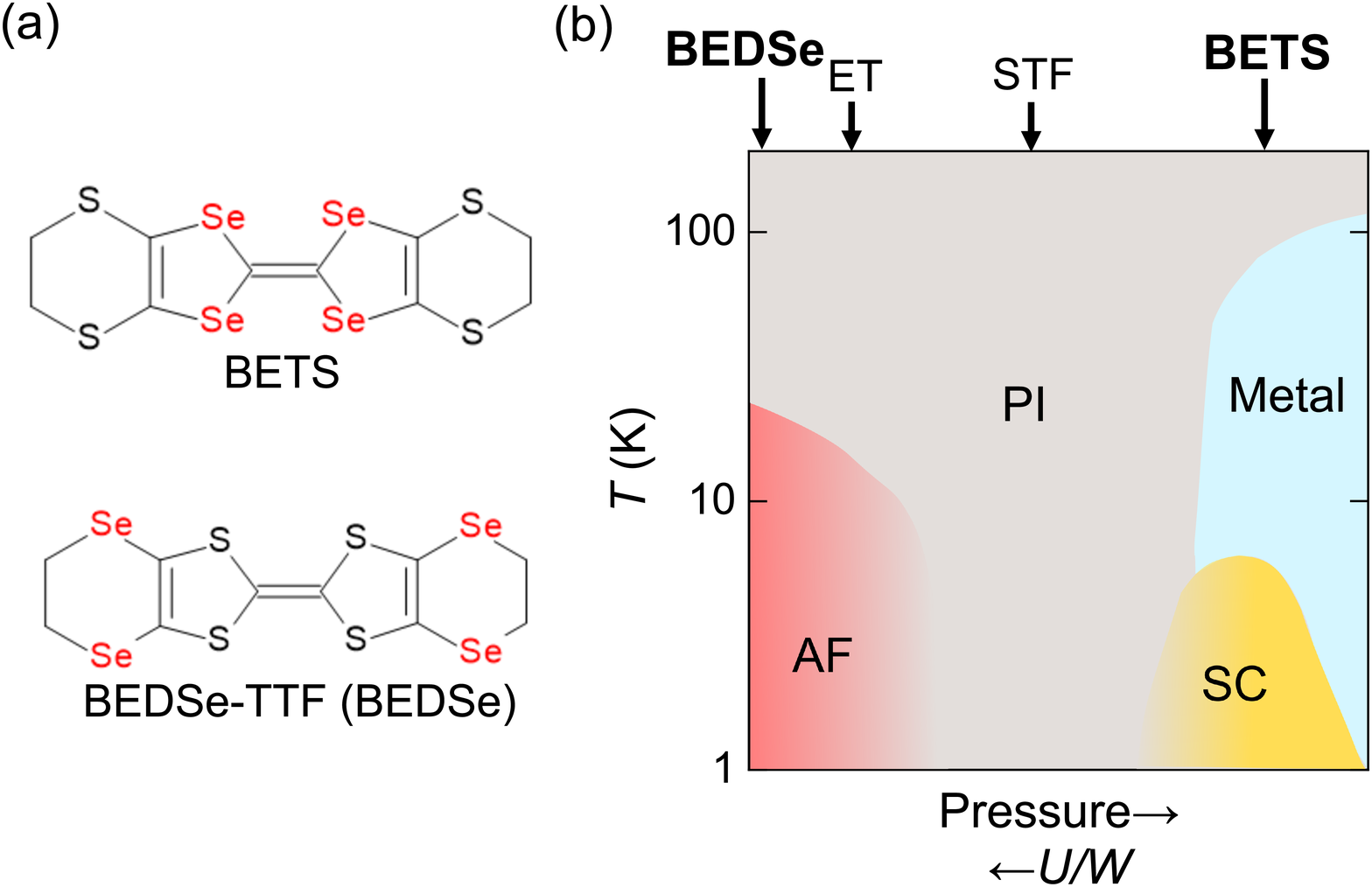}
\end{center}
\caption{(a) Molecular structures of BETS and BEDSe-TTF. 
(b) Temperature-pressure phase diagram of $\lambda$-$D_2$GaCl$_4$ \cite{Ito2022,Mori2001}. 
AF, PI, and SC denote antiferromagnetic, paramagnetic insulating, and superconducting phases, respectively. }
\label{Fig1}
\end{figure}

Interlayer interactions not only act between the conducting layers but also act between the conducting and insulating layers when magnetic ions are introduced into the insulating layer. 
\lFe\ (BETS represents bis(ethylenedithio)tetraselenafulvalene [Fig.~\ref{Fig1}(a)]) exhibits a field-induced superconductivity above $17$~T by the exchange interaction between the conductive $\pi$ electron and the localized $3d$ spin of Fe$^{3+}$ \cite{Uji2001,Balicas2001,Uji2006b}, which is known as $\pi$--$d$ interaction. 
This interaction gives rise to the metal--insulator transition accompanied by AF ordering at zero magnetic fields \cite{Kobayashi1993a,Tokumoto1997}. 
In addition, $\pi$--$d$ interaction causes unusual multistep magnetization processes in $\lambda$-(STF)$_2$FeCl$_4$ and $\lambda$-(BEDSe-TTF)$_2$FeCl$_4$ \cite{Fukuoka2018, Saito2022}, where STF and BEDSe-TTF (noted BEST in Ref.~\cite{Saito2022}) represent unsymmetrical-bis(ethylenedithio)diselenadithiafulvalene and bis(ethylenediseleno)tetrathiafulvalene [Fig.~\ref{Fig1}(a)], respectively. 
These behaviors suggest that interlayer interactions play a crucial role for the various phenomena appearing in Q2D organic systems. 
However, there are few experimental approaches to evaluate the strength or path of the interlayer interaction. 

Meanwhile, in ET-based organic conductors, the dynamics of the ethylene end groups on donor molecules affect the electronic states.
In $\kappa$-type organic conductors, it has been proposed that 
a bad metal state is realized at high temperatures because of its dynamics \cite{Kuwata2011,Matsumoto2014}.
Particularly in $\kappa$-(ET)$_2$Cu[N(CN)$_2$]I, the conformation of the ethylene groups changes the ground state \cite{Naito2022a,Kobayashi2019}. 
Moreover, hydrogen bonding between protons of the ethylene groups and the anion has been discussed \cite{Alemany2012,Pouget2018}; therefore, the relationships between the dynamics of ethylene groups and the interlayer interactions should be addressed. 

To investigate the interlayer interactions and dynamics of ethylene groups (ethylene motion), nuclear magnetic resonance (NMR) spectroscopy using nuclei in anion layers, ``anion NMR'', is suggested. 
Recently, we conducted $^{69,71}$Ga-NMR measurements on a superconductor, $\lambda$-(BETS)$_2$GaCl$_4$ (hereafter \lBETS) \cite{Kobayashi2020b}. 
In the high-temperature region, quadrupolar relaxation, derived from the translational motion of GaCl$_4^-$, was observed. 
We suggested that this molecular motion could be induced by the ethylene motion; however, the study from a structural viewpoint is needed to prove this. 
At low temperatures where the molecular motion freezes, magnetic fluctuation of the electronic system was observed via the transferred hyperfine (HF) interaction between the $\pi$ spins and Ga nuclei. 
This fact suggests that anion NMR is useful for evaluating interlayer interactions. 

While $\lambda$-type salts exhibit physical properties derived from interlayer interactions as seen in $\lambda$-$D_2$FeCl$_4$ ($D$: donor molecule), their physical properties change by donor molecular substitutions due to the electron correlation effect.
In $\lambda$-$D_2$GaCl$_4$, $D$ = ET and BEDSe-TTF salts 
are antiferromagnetic insulators \cite{Saito2018a,Ito2022}.
$D$ = STF salt is a paramagnetic insulator without long-range ordering \cite{Saito2019} and exhibits superconductivity under pressure \cite{Minamidate2015}. 
While $D$ = BETS salt exhibits semiconducting behavior above $100$~K, it becomes a metal and a superconductor below $\sim 6$~K \cite{Kobayashi1995a}. These properties can be understood by the universal phase diagram shown in Fig.~\ref{Fig1} (b) \cite{Ito2022,Mori2001}.
Note that $D$ = BETS and other salts exhibit the same semiconducting behavior near room temperature, but the electrical resistivities of them significantly differ at room temperature, e.g., $0.03$~$\Omega$~cm for $D$ = BETS \cite{Tanaka1999} and $10$~$\Omega$~cm for $D$ = BEDSe-TTF \cite{Cui2005a}. The difference in $U/W$ can explain their electrical conductivity \cite{Ito2022}, where $U$ and $W$ are on-site Coulomb repulsion and bandwidth, respectively. 
How molecular substitutions change the interlayer interactions along with the electronic correlations is also important for a comprehensive understanding of $\lambda$-type salts, including $\lambda$-$D_2$FeCl$_4$.
To investigate this, $^{69,71}$Ga NMR is an effective probe because it can observe the magnetism of $\lambda$-$D_2$GaCl$_4$ from the same Ga site regardless of donor molecule.

In this study, we report $^{69,71}$Ga-NMR measurements on a AF insulator $\lambda$-(BEDSe-TTF)$_2$GaCl$_4$ (hereafter \lBEDSe) and compare the HF coupling constant of \lBEDSe\ with that of a superconductor \lBETS. 
We can discuss the path of the interlayer interaction since S and Se atoms in the BETS molecule are exchanged in the BEDSe-TTF molecule, as shown in Fig.~\ref{Fig1}(a).
In addition, x-ray diffraction measurements were conducted to discuss the relationship between the quadrupolar relaxation at high temperatures and the ethylene motion. 

\section{Experiments}
Single crystals of \lBEDSe\ were synthesized electrochemically \cite{Ito2022}. 
NMR measurements of $^{69}$Ga (nuclear spin $I=\frac{3}{2}$, gyromagnetic ratio $^{69}\gamma/2\pi = 10.219$~MHz/T, nuclear quadrupole moment $^{69}Q = 0.171$~barns) and $^{71}$Ga ($I=\frac{3}{2}$, $^{71}\gamma/2\pi = 12.984$~MHz/T, $^{71}Q = 0.107$~barns) were performed on the central ($\frac{1}{2} \leftrightarrow -\frac{1}{2}$) transition under the magnetic fields of $6.083$~T. 
We used a moderately crushed polycrystalline sample of about $20$~mg, which was randomly oriented with respect to the magnetic field. 
The spectra were obtained via fast Fourier transformation of the echo signal with a $\frac{\pi}{2}$--$\pi$ pulse sequence, where the $\frac{\pi}{2}$ pulse length was typically $3$~$\mu$s. 
The spin-lattice relaxation time, $T_1$, was measured using the conventional saturation-recovery method. 
Single-crystal x-ray diffraction data of \lBEDSe\ were collected using a Bruker SMART APEX II ULTRA diffractometer with Mo-K$\alpha$ radiation ($\lambda = 0.71073$~\AA) at the Comprehensive Analysis Center for Science, Saitama University. 
The crystal structures were solved and refined by SHELXT \cite{shelxt} and SHELXL \cite{shelxl}, respectively. 

\section{Results and Discussion}
\subsection{Spectra}
\label{disA}
\begin{figure}[tbp]
\begin{center}
\includegraphics[width=8.5cm]{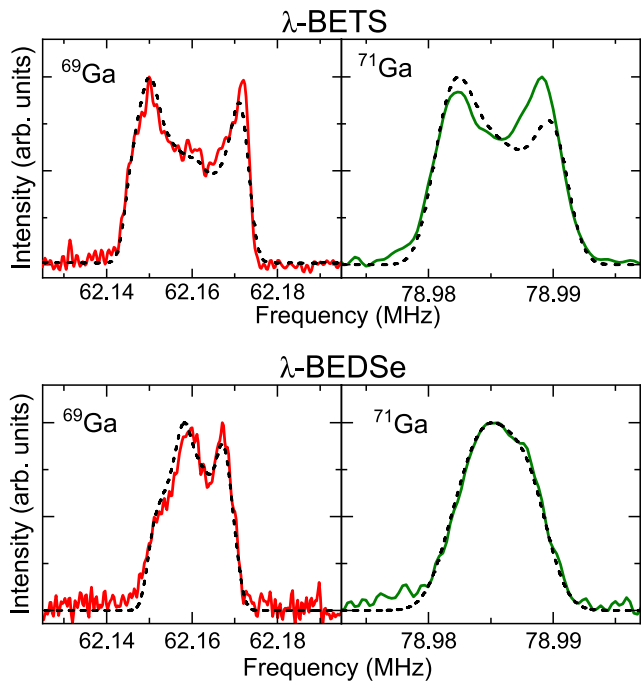}
\end{center}
\caption{$^{69,71}$Ga-NMR central spectra of \lBEDSe\ and \lBETS\ \cite{Kobayashi2020b} at $210$~K. 
Dashed lines are the calculated spectra with parameters listed in Table~\ref{SPsim}.}
\label{SP210K}
\end{figure}

Figure~\ref{SP210K} shows $^{69,71}$Ga-NMR spectra of \lBEDSe\ and \lBETS\ \cite{Kobayashi2020b} at $210$~K originating from the central ($\frac{1}{2}\leftrightarrow-\frac{1}{2}$) transition. 
The spectra with $I=\frac{3}{2}$ can be described by the nuclear spin Hamiltonian as follows:
\begin{align}
\mathcal{H} &= \mathcal{H}_{\rm Z} + \mathcal{H}_{\rm Q} \nonumber \\
 &= -^{n}\gamma\hbar {\bm H} \cdot {\bm I} + \frac{\hbar {}^{n}\omega_{\rm Q}}{6}\left[3I_z^2-{\bm I}^2+\frac{\eta}{2}(I_+^2+I_-^2)\right], 
\end{align}
where $n=69, 71$. 
$\mathcal{H}_{\rm Z}$ is the Zeeman interaction: $\hbar$ and $\bm H$ are the reduced Planck constant and external magnetic field, respectively. 
$\mathcal{H}_{\rm Q}$ is the quadrupolar interaction: $\eta$ and $\omega_{\rm Q}$ denotes the asymmetry parameter of electric field gradient (EFG) and the nuclear quadrupolar frequency, respectively. 
${}^{n}\omega_{\rm Q}$ is defined as ${}^{n}\omega_{\rm Q} = e\,^{n}QV_{ZZ}\,/\,2\hbar$, where $e$ and $V_{ZZ}$ are the elementary charge and principal axis of the EFG, respectively. 
The EFG at the Ga site is almost zero because the Ga nucleus in these materials is tetrahedrally coordinated by four Cl$^{-}$ ions. 
Thus, $\mathcal{H}_{\rm Q}$ is sufficiently smaller than $\mathcal{H}_{\rm Z}$. 
Two peaks were observed in $^{69,71}$Ga-NMR spectra of \lBETS, and the spectral shapes are determined by the powder pattern due to the second-order perturbation effect when $\mathcal{H}_{\rm Q}$ is treated as a perturbation to $\mathcal{H}_{\rm Z}$ \cite{Kobayashi2020b,Abragam1961}. 
The slightly distorted tetrahedral coordination of GaCl$_4^-$ causes the finite EFG at the Ga site. 

\begin{table}[tbp]
\caption{
Parameters of the calculated $^{69,71}$Ga-NMR spectra depicted by dashed lines in Fig.~\ref{SP210K}. 
}
\begin{ruledtabular}
\begin{tabular}{ccc}
& \lBEDSe & \lBETS \\ 
\hline 
${}^{69}\omega_{\rm Q}/2\pi$ (MHz) & $1.35$ & $1.73$ \\
${}^{71}\omega_{\rm Q}/2\pi$ (MHz) & $0.94$ & $1.20$ \\
$\eta$ & $0.43$ & $0.17$ \\
\end{tabular}
\end{ruledtabular}
\label{SPsim}
\end{table}

Although the Ga site of \lBEDSe\ is in the same situation as that of \lBETS, spectral splitting due to second-order perturbation was observed only in $^{69}$Ga NMR, implying that the EFG at the Ga site in \lBEDSe\ is smaller than that in \lBETS\ because the splitting interval is proportional to $\omega_{\rm Q}^2$ \cite{Cohen1957}. 
To discuss the difference in the spectra of the two salts quantitatively, we carried out numerical simulation to reproduce them, as shown by the dashed lines in Fig.~\ref{SP210K}. 
$^{69,71}$Ga-NMR spectra of each salt can be reasonably reproduced by the parameters shown in Table~\ref{SPsim}, where ${}^{n}\omega_{\rm Q} \propto {}^{n}Q$. 
In terms of the results of \lBETS, the parameters are consistent with those obtained from simulation of the overall $^{69}$Ga-NMR spectrum including satellite peaks at $80$~K \cite{Kobayashi2020b}. 
When the results of both salts are compared, ${}^{n}\omega_{\rm Q}$ of \lBEDSe\ is smaller than that of \lBETS, resulting in no splitting in the $^{71}$Ga-NMR spectra with smaller $Q$. 
Reference~\cite{Schurko2002} suggested that the EFG tensor at the tetrahedrally coordinated central atom is sensitive to the arrangement of the surrounding ions. 
The difference in ${}^{n}\omega_{\rm Q}$ between \lBETS\ and \lBEDSe\ is only $\sim 20$~\% despite the sensitivity to Cl$^-$ ion arrangement, suggesting that the GaCl$_4^-$ in both salts is almost in the same environment.

\begin{figure}[tbp]
\begin{center}
\includegraphics[width=8cm]{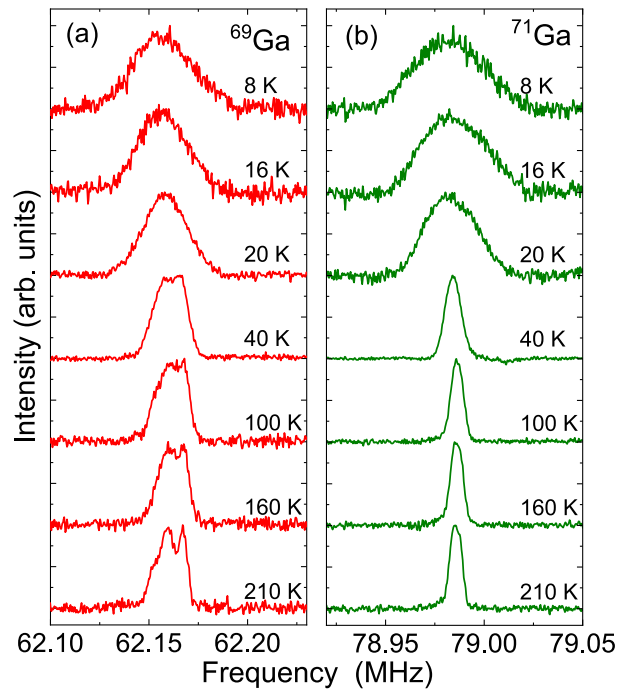}
\end{center}
\caption{Central spectra of (a) $^{69}$Ga and (b) $^{71}$Ga NMR of \lBEDSe\ at various temperatures. }
\label{SP}
\end{figure}

Figure~\ref{SP} shows the temperature evolution of $^{69,71}$Ga-NMR spectra of \lBEDSe. 
Whereas the spectral shapes are almost unchanged above $40$~K, significant line broadening was observed below $20$~K. 
This behavior is understood as the development of the internal magnetic field due to the AF ordering of the $\pi$-spin system at $T_{\rm N} = 22$~K \cite{Ito2022}. 
This observation suggests that $^{69,71}$Ga-NMR spectra can detect the static magnetic properties of the $\pi$ spin system of \lBEDSe, although the Ga site is far from the $\pi$ spins. 

\begin{figure}[tbp]
\begin{center}
\includegraphics[width=8cm]{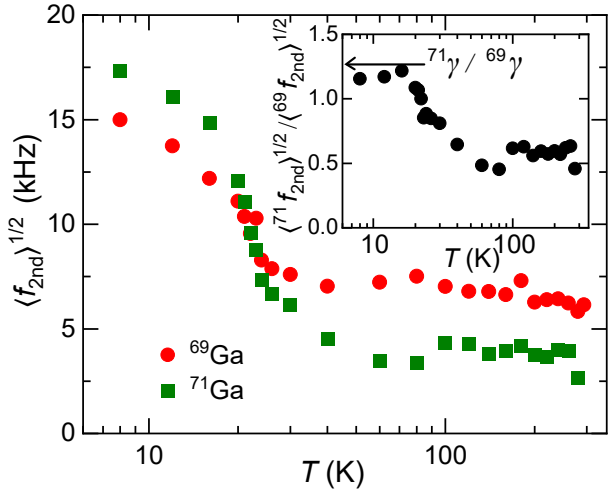}
\end{center}
\caption{Second moment $\langle ^{69,71}f_{\rm 2nd}\rangle^{1/2}$ of \lBEDSe\ as a function of temperature. 
Inset shows the temperature dependence of the isotopic ratio of $\langle f_{\rm 2nd}\rangle^{1/2}$. 
The arrow represents the value of $^{71}\gamma\,/\,^{69}\gamma$~. }
\label{secondmoment}
\end{figure}

Gaussian-shaped spectra were observed below $20$~K, whereas the spectral shapes above $40$~K are dominated by the powder pattern due to the second-order perturbation. 
To phenomenologically evaluate the linewidth, we estimated the square root of the second moment: $\langle f_{\rm 2nd}\rangle^{1/2} = \left[\int I(f)\:(f-\langle f\rangle)^2\:df/\int I(f)\:df\right]^{1/2}$, where $\langle f \rangle=\int f\:I(f)\:df/\int I(f)\:df$ is the first moment and $I(f)$ is the spectral intensity as a function of frequency $f$. 
Figure~\ref{secondmoment} shows the temperature dependence of $\langle f_{\rm 2nd}\rangle^{1/2}$ of $^{69,71}$Ga-NMR spectra, $\langle^{69,71}f_{\rm 2nd}\rangle^{1/2}$. 
$\langle^{69,71}f_{\rm 2nd}\rangle^{1/2}$ increase sharply below approximately $40$~K owing to the magnetic fluctuation of the $\pi$ spins. 
In addition, $\langle^{69}f_{\rm 2nd}\rangle^{1/2}$ is larger than $\langle^{71}f_{\rm 2nd}\rangle^{1/2}$ at high temperatures; however, the relationship is reversed at low temperatures. 
This indicates that the second moment is dominated by the quadrupolar interaction at high temperatures ($^{69}Q > {}^{71}Q$) and the magnetic interaction at low temperatures ($^{71}\gamma > {}^{69}\gamma$). 
The inset of Fig.~\ref{secondmoment} shows that the isotopic ratio of $\langle f_{\rm 2nd}\rangle^{1/2}$ below $18$~K is almost identical to $^{71}\gamma/^{69}\gamma$. 

In $^{69,71}$Ga-NMR spectral measurements, we observed spectra with the typical shapes generated by the finite EFG at the Ga site in the high-temperature region as well as \lBETS\ and confirmed the line broadening due to the AF ordering of the $\pi$ spin system despite the Ga site being far from the $\pi$ spins.
Thus, the spectral study enable us to discuss the static properties of both the EFG and magnetic field at the Ga site. 

\subsection{Dynamics of molecular motion in high-temperature region}
\label{disB}
$T_1$ measurements allow us to evaluate the magnetic and EFG fluctuations and to distinguish which is dominant using the difference in the properties of the two isotopes. 
The recovery curves for the central ($\frac{1}{2}\leftrightarrow-\frac{1}{2}$) transition were fitted using the following function; $1-M(t)/M(\infty)=0.1\exp[-(t/T_{1})^\beta]+0.9\exp[-(6t/T_{1})^\beta]$, 
where $M(t)$ is the nuclear magnetization at time $t$ after the saturation, $M(\infty)$ is the nuclear magnetization at equilibrium ($t\rightarrow\infty$), and $\beta$ is the stretch exponent. 
The recovery curves above $100$~K were well fitted by the function with $\beta = 1$, whereas $\beta$ decreased towards $T_{\rm N}$ at low temperatures as discussed in Sec.~\ref{disC}.

\begin{figure}[tbp]
\begin{center}
\includegraphics[width=8cm]{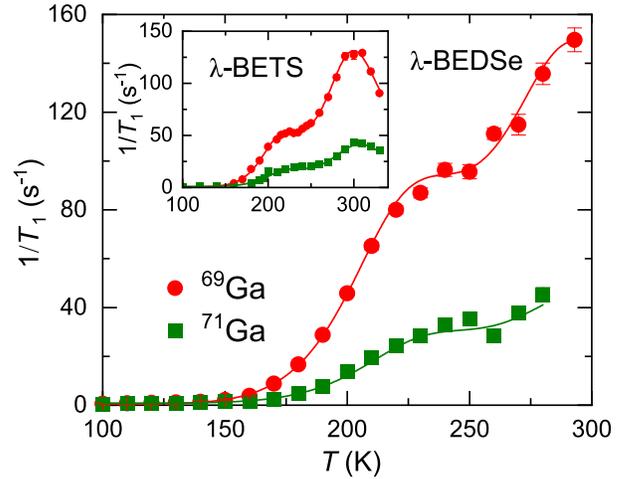}
\end{center}
\caption{$^{69,71}T_{1}^{-1}$ of \lBEDSe\ above $100$~K as a function of temperature. 
Inset shows $^{69,71}T_{1}^{-1}$ of \lBETS\ \cite{Kobayashi2020b}.
Solid lines are the fitting curves by Eq.~(\ref{BPP}) with the fitting parameters listed in Table~\ref{para}. }
\label{T1H}
\end{figure}

\begin{table}[tbp]
\caption{Parameters for $^{69}T_{1}^{-1}$ above $100$~K fitted by Eq.~(\ref{BPP}). }
\begin{ruledtabular}
\begin{tabular}{ccccc}
& \multicolumn{2}{c}{\lBEDSe} & \multicolumn{2}{c}{\lBETS} \\
& $i=1$ & $i=2$ & $i=1$ & $i=2$ \\ 
\hline 
$E_{{\rm A},i}/k_{\rm B}$ ($10^3$~K) & $2.0(1)$ & $3.2(7)$ & $1.9(1)$ & $2.9(2)$ \\
$\tau_{0,i}$ ($10^{-13}$~s) & $4.3(16)$ & $0.5(14)$ & $3.8(17)$ & $1.6(9)$ \\
$\langle ^{69}\omega_{{\rm Q},i}^2 \rangle^{1/2}/2\pi$ (kHz) & $281(12)$ & $353(8)$ & $215(4)$ & $348(4)$ \\
$\langle ^{71}\omega_{{\rm Q},i}^2 \rangle^{1/2}/2\pi$ (kHz) & $176(7)$ & $220(5)$ & $134(2)$ & $218(2)$ \\
$^{69}T_{1, \rm{m}}^{-1}$ (s$^{-1}$) & \multicolumn{2}{c}{$0.52(7)$} & \multicolumn{2}{c}{$0.61(9)$} \\
$^{71}T_{1, \rm{m}}^{-1}$ (s$^{-1}$) & \multicolumn{2}{c}{$0.84(12)$} & \multicolumn{2}{c}{$0.98(14)$} \\
\end{tabular}
\end{ruledtabular}
\label{para}
\end{table}

Figure~\ref{T1H} shows the temperature dependence of $^{69,71}T_{1}^{-1}$ ($T_1^{-1}$ of $^{69,71}$Ga NMR) of \lBEDSe\ above $100$~K. 
$^{69}T_{1}^{-1}$ is greater than $^{71}T_{1}^{-1}$ in the high-temperature range and the ratio corresponds to the ratio of $Q^2$, indicating that the quadrupolar relaxation mechanism is dominant. 
Above $150$~K, $^{69,71}T_{1}^{-1}$ strongly depend on temperature and exhibit shoulder-like anomalies at around $230$~K. 
Similar behavior was also observed in $^{69,71}T_1^{-1}$ of \lBETS\ (inset of Fig.~\ref{T1H}) \cite{Kobayashi2020b}, which was explained by the Bloembergen-Purcell-Pound (BPP) formula \cite{Bloembergen1948}. 
We conducted the same analysis with slight modification \cite{BPPfit} for $^{69,71}T_1^{-1}$ of \lBEDSe\ and \lBETS. 
$^{n}T_{1}^{-1}$ ($n=69,71$) with $I=\frac{3}{2}$ can be written as \cite{Abragam1961,Spiess1978,Avogadro1990}, 
\begin{align}
\frac{1}{^{n}T_{1}}=\sum_{i=1,2}\frac{\langle^{n}\omega_{{\rm Q},i}^2\rangle}{50} \frac{\tau_{{\rm c},i}}{1+{}^{n}\omega_{\rm L}^{2}\tau_{{\rm c},i}^{2}}+\frac{1}{^{n}T_{1,{\rm m}}}.
\label{BPP}
\end{align}
$^{n}\omega_{\rm L}$ is the Larmor frequency ($^{69}\omega_{\rm L}/2\pi=62.16$~MHz and $^{71}\omega_{\rm L}/2\pi=78.99$~MHz).
An effective root-mean-square quadrupole coupling frequency, $\langle^{n}\omega^{2}_{{\rm Q},i}\rangle^{1/2}$, originates from the modulation of EFG at different equilibrium positions. 
Since it is proportional to $^{n}Q$, the isotopic ratio of $\langle^{69}\omega^{2}_{{\rm Q},i}\rangle^{1/2}$ to $\langle^{71}\omega^{2}_{{\rm Q},i}\rangle^{1/2}$ is fixed to $^{69}Q/^{71}Q$. 
$\tau_{{\rm c},i}$ is the correlation time described by Arrhenius-type temperature dependence, $\tau_{{\rm c},i}=\tau_{0,i}\exp(E_{{\rm A},i}/k_{\rm B}T)$, with a prefactor, $\tau_{0,i}$, activation energy of molecular motion, $E_{{\rm A},i}$, and Boltzman constant, $k_{\rm B}$. 
$^{n}T_{1,\rm m}^{-1}$ is a parameter that originates from the magnetic fluctuation of the $\pi$ spins in the low-temperature region, where we assumed the constant $^{n}T_{1,\rm m}^{-1}$, and the ratio $^{71}T_{1,\rm m}^{-1}/^{69}T_{1,\rm m}^{-1} = (^{71}\gamma/^{69}\gamma)^2$. 
As represented by the solid lines in Fig.~\ref{T1H}, the fitting by Eq.~(\ref{BPP}) reproduces the temperature dependence of $^{69,71}T_1^{-1}$ well, including the reported results of \lBETS\ \cite{Kobayashi2020b}. 
The obtained fitting parameters for both salts shown in Table~\ref{para} are similar each other, indicating that the quadrupolar relaxation in the high-temperature region can be explained by the same mechanism. 
This mechanism has been interpreted as the translational motion of GaCl$_4^{-}$ ions, resulting in the EFG fluctuation at the Ga site \cite{Kobayashi2020b}. 
In addition, the nearly identical $E_{{\rm A},i}$ and $\tau_{0,i}$ between \lBETS\ and \lBEDSe\ show that the donor molecular substitution effect on the time scale of the molecular motion is negligible. 

\begin{figure}[tbp]
\begin{center}
\includegraphics[width=6cm]{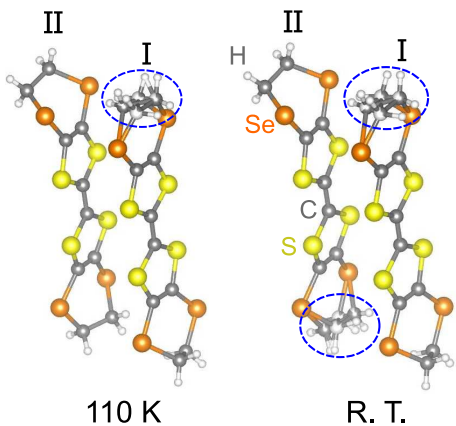}
\end{center}
\caption{
Donor molecular structures of \lBEDSe\ at room temperature (R.T.) and $110$~K. 
The area surrounded by round dashed lines represents the disordered ethylene groups.
}
\label{disorder}
\end{figure}

In Eq.~(\ref{BPP}), we assumed two components of the BPP formula ($i=1,2$), as shown in Table \ref{para}. 
The components of $i=1$ and $2$ correspond to the shoulder-like structures at around $230$~K and the further enhancement above $250$~K of $^{69,71}T_1^{-1}$ (the latter corresponds to the peak at $300$~K in $^{69,71}T_1^{-1}$ of \lBETS\ shown in the inset of Fig.~\ref{T1H} \cite{Kobayashi2020b}), respectively. 
These results indicate that there are two types of molecular motion with different parameters. 
Moreover, the translational motion of GaCl$_4^-$ can be induced by the conformational motion of ethylene groups near anion layers, as mentioned in the previous $^{69,71}$Ga-NMR study on \lBETS\ \cite{Kobayashi2020b}. 
To further clarify the relationship between the observed $^{69,71}T_1^{-1}$ and the ethylene motion, x-ray structural analysis was performed at room temperature ($293$~K) and $110$~K \cite{xray}. 
Figure~\ref{disorder} shows the crystallographically inequivalent molecules I and II extracted from the crystal structure of \lBEDSe. 
At room temperature, the ethylene groups at one end of both molecules I and II are disordered, occupying staggered and eclipsed conformations with a ratio of occupancy $0.636(6):0.364(6)$ (molecule I) and $0.131(11):0.869(11)$ (molecule II). 
At $110$~K, molecule I still has disordered sites with the occupancy of $0.797(5):0.203(5)$, whereas molecule II is completely ordered in the eclipsed conformation. 
Therefore, the ethylene motion of molecule I that remains disordered at lower temperatures is assigned to the $i=1$ component in Eq.~(\ref{BPP}), contributing to quadrupolar relaxation on the low-temperature side, and that of molecule II is assigned to $i=2$ component. 
This relationship between the two components in the quadrupolar relaxation and the ethylene groups on the two crystallographically independent molecules is considered to hold also in \lBETS\ \cite{Kobayashi2020b}. 

In addition to the fact that \lBETS\ has two types of the ethylene motion \cite{Kobayashi2020b}, the x-ray structural analyses of $\lambda$-(BEDSe-TTF)$_2$FeCl$_4$ show that the ethylene groups at the same sites are disordered \cite{Saito2022}, implying that a similar degree of ethylene motion commonly exists in the $\lambda$-type salts. 
These results will aid our understanding of the mechanisms that cause $\pi$--$d$ spin correlations and dielectric anomalies that develop at low temperatures where ethylene motion freezes, as reported for $\lambda$-(BETS)$_2$FeCl$_4$ \cite{Lee2018,Matsui2001}. 

\subsection{Hyperfine interaction between Ga nuclear spin and $\pi$ spin}
\label{disC}
\begin{figure}[tbp]
\begin{center}
\includegraphics[width=8.5cm]{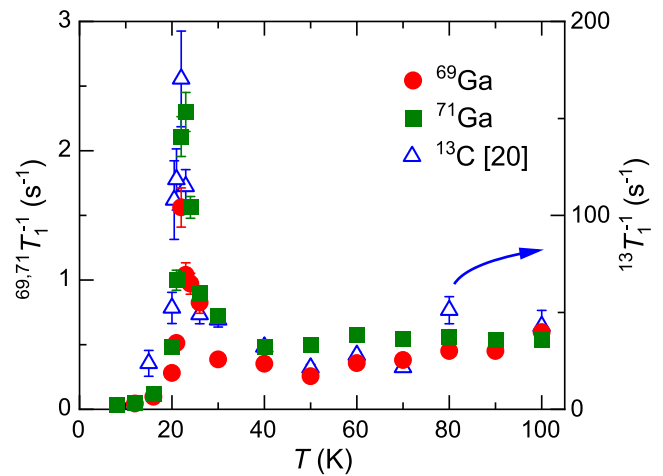}
\end{center}
\caption{
$^{69,71}T_1^{-1}$ (left axis) and  $^{13}T_1^{-1}$ (right axis) \cite{Ito2022} of \lBEDSe\ as a function of temperature. 
} 
\label{T1L}
\end{figure}

\begin{figure}[tbp]
\begin{center}
\includegraphics[width=8.5cm]{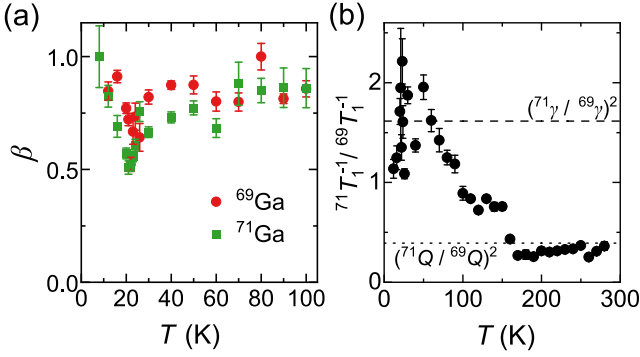}
\end{center}
\caption{
(a) Stretch exponent $\beta$ of $^{69,71}$Ga NMR below $100$~K and (b) $^{71}T_1^{-1}/^{69}T_1^{-1}$ of \lBEDSe\  as functions of temperature. 
Dashed and dotted lines represent the isotopic ratio of $\gamma^2$ and $Q^2$, respectively. 
} 
\label{fig7}
\end{figure}

Figure~\ref{T1L} shows the temperature dependence of $^{69,71}T_{1}^{-1}$ of \lBEDSe\ below $100$~K.
Above $40$~K, $^{69,71}T_{1}^{-1}$ are independent of temperature, but below $40$~K, they rapidly increase and show the divergent behavior toward $T_{\rm N}=22$~K. 
The divergence of $^{69,71}T_{1}^{-1}$ can be attributed to the magnetic transition since the line broadening around $T_{\rm N}$ is also magnetic, as shown in Fig.~\ref{secondmoment}. 
In addition, the stretch exponent, $\beta$, depends on temperatures below $100$~K, as shown in Fig.~\ref{fig7}(a).
With a decrease in temperature, $\beta$ decreases from $1$ to a minimum at $T_{\rm N} = 22$~K. 
$\beta$ approaches $1$ as the temperature further decreases. 

The isotopic ratio of $T_{1}^{-1}$, $^{71}T_{1}^{-1}/{}^{69}T_{1}^{-1}$, also confirms that the low-temperature relaxation is due to magnetic fluctuations. 
Figure~\ref{fig7}(b) shows the temperature dependence of $^{71}T_{1}^{-1}/{}^{69}T_{1}^{-1}$. 
Above $150$~K, $^{71}T_{1}^{-1}/{}^{69}T_{1}^{-1}$ values coincide with $\left(^{71}Q/{}^{69}Q\right)^2$, as mentioned in Sec.~\ref{disB}, depicting the dominance of EFG fluctuation. 
Below $150$~K, $^{71}T_{1}^{-1}/{}^{69}T_{1}^{-1}$ deviates from $\left(^{71}Q/{}^{69}Q\right)^2$ and approaches $\left(^{71}\gamma/{}^{69}\gamma\right)^2$, indicating that the magnetic fluctuation becomes dominant as the temperature decreases. 

The observed magnetic fluctuation originates from the spin fluctuation of the $\pi$ spins in the donor layer since the Ga ion is nonmagnetic. 
The temperature dependence of $T_{1}^{-1}$ of $^{13}$C NMR, $^{13}T_{1}^{-1}$, is also plotted in Fig.~\ref{T1L} for comparison \cite{Ito2022}. 
Both $^{69,71}T_{1}^{-1}$ and $^{13}T_{1}^{-1}$ qualitatively have the same temperature dependence. 
Thus, the magnetic fluctuation of the $\pi$-spin system was probed by $^{69,71}$Ga NMR in \lBEDSe\ as well as in \lBETS\ \cite{Kobayashi2020b}. 
These magnetic fluctuations were observed by the suppression of EFG fluctuations as the ethylene motion freezes. 

\begin{figure}[tbp]
\begin{center}
\includegraphics[width=8cm]{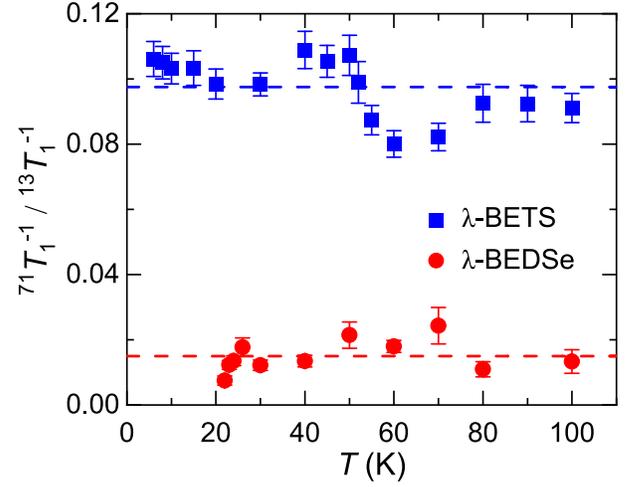}
\end{center}
\caption{
Temperature dependence of $^{71}T_{1}^{-1}/{}^{13}T_{1}^{-1}$ of \lBETS\ and that of \lBEDSe\ in the region where the spin fluctuation is dominant. 
They are obtained from the present and reported data  \cite{Kobayashi2020b, Kobayashi2017, Ito2022}. 
Dashed lines represent the average values of $^{71}T_{1}^{-1}/^{13}T_{1}^{-1}$ of both salts. 
} 
\label{T1ratio}
\end{figure}

When the magnetic relaxation is due to the spin fluctuation of an electronic system, $T_{1}^{-1}$ is generally written as follows \cite{Moriya1963}:
\begin{align}
\label{spinlattice}
\frac{1}{^{n}T_1}=\frac{2\:^{n}\gamma^{2}k_{\rm B}T}{\gamma_{\rm e}^{2}\hbar^{2}}\sum_{\bm{q}}|^{n}A_{\bm{q}}|^{2}\frac{\chi''(\bm{q})}{^{n}\omega_{\rm L}},
\end{align}
where $\gamma_{\rm e}$ is the gyromagnetic ratio of the electron, $^{n}A_{\bm{q}}$ is the wave vector $\bm q$-dependent HF coupling constant, and $\chi''(\bm q)$ is the imaginary part of the dynamic susceptibility. 
Ga sites feel the transferred HF and dipole fields from the donor sites.
First, the dipole field from the donor molecules can be calculated electromagnetically \cite{dipole} and is of the order of several $10$~Oe/$\mu_{\rm B}$. 
This estimate can be confirmed from $^{69,71}$Ga-NMR spectra in the AF state.
Because the external field in the present experiment is larger than the spin-flop field as expected from the typical ET-based antiferromagnet \cite{Taniguchi2005}, the line broadening of $^{69,71}$Ga NMR is dominated by the dipole field.
As can be seen in $\langle^{71}f_{\rm 2nd}\rangle^{1/2}$ (Fig.~\ref{secondmoment}), the line broadening by AF ordering is $\sim 14$~kHz, which corresponds to $11$~Oe. 
This is consistent with the calculated value.
Next, we compare the calculated HF interaction from the dipole field with experiments.
For simplicity, the following discussion uses the results for ${}^{71}T_1^{-1}$ with larger $\gamma$. 
In ${}^{71}T_1^{-1}$ and ${}^{13}T_1^{-1}$, the $\sum_{\bm q}\chi''(\bm{q})$ part in Eq.~(\ref{spinlattice}) should be the same since the same magnetic fluctuation appeared. 
Therefore, 
\begin{equation}
    \frac{{}^{71}T_1^{-1}}{{}^{13}T_1^{-1}} = \frac{{}^{71}\gamma^2 \sum_{\bm q}|^{71}A_{\bm{q}}|^{2}/{}^{71}\omega_{\rm L}}{{}^{13}\gamma^2 \sum_{\bm q}|^{13}A_{\bm{q}}|^{2}/{}^{13}\omega_{\rm L}}.
    \label{ratio7113}
\end{equation}
Figure \ref{T1ratio} shows $^{71}T_{1}^{-1}/{}^{13}T_{1}^{-1}$ below $100$~K in the paramagnetic state, where the magnetic fluctuations dominate.
As depicted by the dashed line, an average value of $^{71}T_{1}^{-1}/{}^{13}T_{1}^{-1}$ is $0.015(5)$ in \lBEDSe, resulting in $( \sum_{\bm q}|^{71}A_{\bm{q}}|^{2}/\sum_{\bm q}|^{13}A_{\bm{q}}|^{2} ) ^{1/2}$ = $0.11(2)$.
$^{13}A$ is approximately $\sim 5$~kOe/$\mu_{\rm B}$, referring to the \lBETS\ results \cite{Kobayashi2017}.
Applying the same magnitude of $^{13}A$ to \lBEDSe\ (details below) and neglecting the $\bm q$ dependence of the HF coupling constant, the observed HF coupling constant of $^{71}$Ga NMR is estimated to be $5$~kOe/$\mu_{\rm B} \times 0.11 = 550$~Oe/$\mu_{\rm B}$.
This is an order of magnitude larger than the value of the dipole field, indicating that the transferred HF interaction is the dominant as the HF mechanism.

We now discuss the magnitude of the transferred HF interaction for \lBETS~and \lBEDSe\ from the ratio of the values of Eq.~(\ref{ratio7113}).
All the data used in the following discussion were measured at the same condition, and values other than the HF coupling constant cancel out.
We obtained the following equation:
\begin{align}
&\frac{({}^{71}T_{1}^{-1}/{}^{13}T_{1}^{-1})_{\rm BETS}}{({}^{71}T_{1}^{-1}/{}^{13}T_{1}^{-1})_{\rm BEDSe}}& \nonumber\\
&\approx\frac{({}^{71}A/{}^{13}A)_{\rm BETS}^{2}}{({}^{71}A/{}^{13}A)_{\rm BEDSe}^{2}} \approx\frac{{}^{71}A_{\rm BETS}^{2}}{{}^{71}A_{\rm BEDSe}^{2}}, 
\label{ratiodonor}
\end{align}
where the donor molecules are indicated as subscripts. 
The donor molecular substitution effect on ${}^{13}A$ was ignored in the latter approximation since $\lambda$-$D_2$GaCl${}_4$ are isostructural and the local environments around ${}^{13}$C nucleus are similar. 
Actually, ${}^{13}A_{\rm STF}$ and ${}^{13}A_{\rm BETS}$ are of the same order \cite{Saito2019,Kobayashi2017}, where ${}^{13}A_{\rm STF}$ is the HF coupling constant in $\lambda$-(STF)$_2$GaCl$_4$. 
Therefore, Eq.~(\ref{ratiodonor}) allows us to compare the magnitude of the HF interaction between \lBETS\ and \lBEDSe\ from $T_1^{-1}$ measurements. 
As shown in Fig.~\ref{T1ratio}, the average values were obtained as $({}^{71}T_{1}^{-1}/{}^{13}T_{1}^{-1})_{\rm BETS}=0.098(9)$ and $({}^{71}T_{1}^{-1}/{}^{13}T_{1}^{-1})_{\rm BEDSe}=0.015(5)$, resulting in ${}^{71}A_{\rm BETS}/{}^{71}A_{\rm BEDSe}=2.6(4)$. 
This result implies that the magnitude of the transferred HF interaction in \lBETS\ is more than two times larger than that in \lBEDSe. 
The difference in the magnitudes between \lBETS\ and \lBEDSe\ suggests that the proton-Cl contacts are not dominant for the transferred HF interaction. 

The variation in physical properties due to the donor molecular substitution in $\lambda$-$D_2$GaCl$_4$ has been discussed in terms of chemical pressure effect \cite{Mori2001}, with \lBEDSe\ compound located at the most negative position on the pressure axis (Fig.~\ref{Fig1}(b)) \cite{Ito2022}. 
Our findings indicate that replacing BETS or STF molecules with BEDSe-TTF molecule causes negative-pressure effect on not only intralayer but also interlayer interactions, which may be necessary for a comprehensive understanding of their physical properties.

\begin{figure}[tbp]
\begin{center}
\includegraphics[width=8cm]{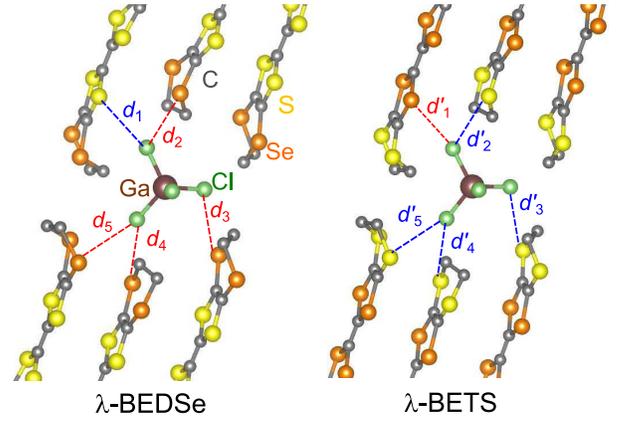}
\end{center}
\caption{
Short contacts between Cl and chalcogens in \lBEDSe\ and \lBETS. 
Blue (red) dashed lines represent short contacts between S (Se) and Cl within the van der Waals distances of $3.6$~\AA\ (S--Cl) and $3.7$~\AA\ (Se--Cl): \lBEDSe; $d_1=3.5978(13)$~\AA, $d_2=3.2601(8)$~\AA, $d_3=3.5230(11)$~\AA, $d_4=3.5139(9)$~\AA, $d_5=3.6301(8)$~\AA, \lBETS; $d'_1=3.5144(12)$~\AA, $d'_2=3.4084(12)$~\AA, $d'_3=3.5699(15)$~\AA, $d'_4=3.6006(14)$~\AA, $d'_5=3.5540(12)$~\AA. 
}
\label{path}
\end{figure}

Some Se/S--Cl contacts shorter than van der Waals distances (Se--Cl: $3.7$~\AA, S--Cl: $3.6$~\AA\ \cite{Batsanov2001}) exist in \lBETS\ \cite{Tanaka1999, Ito2022} and \lBEDSe, as shown in Fig.~\ref{path}, which significantly contribute to the transferred HF interaction. 
The short Se--Cl contact would contribute more to the interaction than the S--Cl contact since the atomic radius of Se is larger than that of S. 
Although the number of short Se--Cl contacts in \lBEDSe\ is greater than that in \lBETS\ with an extremely short Se--Cl contact of $d_2=3.2601(8)$~\AA, 
this study demonstrated that the transferred HF interaction of \lBEDSe\ is smaller than that of \lBETS. 
Therefore, the paths through the inner chalcogens and Cl are more essential for the transferred HF interaction. 
The inner chalcogens in the ET molecule have larger coefficients of the highest occupied molecular orbital proportional to the electron density than the outer chalcogens \cite{Mori1984}, which could be why the paths through the inner chalcogen and Cl are crucial. 

\vspace{1cm}

As discussed above, the coupling constants between $\pi$ electrons and the Ga nuclear spins were obtained by combining the results of $^{69,71}$Ga-NMR and $^{13}$C-NMR measurements. 
This enables us to explain the coupling between the conduction layers via the anion layer although such a discussion has been difficult until now. 
In addition to the difference in the electronic states between \lBEDSe\ and \lBETS, this study reveals that the interlayer interaction is smaller in \lBEDSe. 
Thus, we demonstrate that anion NMR experiment is a powerful method for investigating the interlayer interactions. 
The contribution of such interactions has also been suggested for $\kappa$-type organic antiferromagnets. 
Since the $\kappa$-(ET)$_2$Cu[N(CN)$_2$]Cl and deuterated $\kappa$-(ET)$_2$Cu[N(CN)$_2$]Br salts are isostructural, the same magnetic structure is expected from the anisotropy resulting from dipole interactions. 
However, magnetization measurements and numerical simulations have recently revealed that the magnetic structures are different between them \cite{Ishikawa2018,Oinuma2020}. 
In organic magnets, which are similar to the Heisenberg spin systems, interlayer coupling could be particularly important for three-dimensional order.
Anion NMR will be helpful for clarifying this issue and also the development of interlayer coupling upon cooling \cite{Antal2011, Antal2012a}. 

Finally, we discuss the relationship between the transferred HF interaction in $\lambda$-$D_2$GaCl$_4$ and the $\pi$--$d$ interaction in $\lambda$-$D_2$FeCl$_4$. 
The transferred HF interaction is between the $\pi$ spin and the Ga nuclear spin, whereas the $\pi$--$d$ interaction is between the $\pi$ spin and the $3d$ spin of Fe${}^{3+}$ ion. 
This study experimentally revealed the importance of the path through the inner chalcogen and Cl for the transferred HF interactions. 
Mori and Katsuhara also reported that the interaction via short Se--Cl contacts has the largest contribution to the $\pi$--$d$ interaction in \lFe~from the calculation of the intermolecular overlap integrals \cite{Mori2002}.
Therefore, the transferred HF and $\pi$--$d$ interactions can be attributed to the same mechanism.
The reported $\lambda$-$D_2$FeCl$_4$ salts are $D$ = BETS, STF, and BEDSe-TTF, which are antiferromagnets with $T_{\rm N} = 8.5, 16$, and $25$~K, respectively \cite{Tokumoto1997,Fukuoka2018,Saito2022}. 
Among them, \lFe\ is expected to have the largest $\pi$--$d$ interaction, and \lBEDSeFe\ is expected to have the smallest $\pi$--$d$ interaction due to the weak interaction between inner S and Cl \cite{Saito2022}. 
Magnetic susceptibility of \lBEDSeFe\ does not show an anisotropy \cite{Saito2022} as well as that of $\lambda$-(BETS)$_2$FeBr$_x$Cl$_{4-x}$ for $0.3 < x < 0.5$ \cite{Akutsu1998}, whereas those of \lFe\ and $\lambda$-(STF)$_2$FeCl$_4$ show an anisotropy below $8$~K \cite{Tokumoto1997,Minamidate2018}. 
Akutsu {\it et al.} claimed that the $\pi$--$d$ interaction in $\lambda$-(BETS)$_2$FeBr$_x$Cl$_{4-x}$ for $0.3 < x < 0.5$ is weaker than that in \lFe\ \cite{Akutsu1998}. 
Thus, \lBEDSeFe\ has been referred to as a weaker $\pi$--$d$ system, which is consistent with our findings. 

\section{Summary}
We performed $^{69,71}$Ga-NMR and x-ray diffraction measurements on an organic antiferromagnet, \lBEDSe.
The spectral measurements revealed a powder pattern due to finite EFG at high temperatures and a spectral broadening due to AF ordering at low temperatures, indicating that the static nature of EFG and magnetic fields can be detected even by NMR at Ga sites far from the electron system.
In the $T_1$ measurements, we observed the quadrupolar relaxation described by the BPP formula with two components in the high-temperature region. 
By x-ray structural analysis, it was revealed that the two components were assigned to the two crystallographically independent ethylene groups. 
In the low-temperature region where the ethylene motion freezes, the magnetic fluctuation of the $\pi$-spin system was observed through the transferred HF interaction between the $\pi$ spin and the Ga nuclear spin, as in the case of \lBETS. 
We discovered that \lBETS\ had a lager transferred HF interaction than \lBEDSe. 
This finding enables us to conclude that the short contacts between the chalcogens of the fulvalene part of the donor molecules and the Cl of anion molecules are the most significant contributors to this interaction. 
The transferred HF interaction in $\lambda$-$D_2$GaCl$_4$ and the $\pi$--$d$ interaction in the isostructural $\lambda$-$D_2$FeCl$_4$ can be derived from the same path. 
Therefore, we also evaluated the $\pi$--$d$ interaction in the $\lambda$-type $\pi$--$d$ system. 
Present study reveals that anion NMR measurement is an effective tool for studying the interlayer interaction for which experimental approaches are limited in Q2D organic conductors. 

\section*{Acknowledgments}
We would like to acknowledge R. Saito (Hokkaido Univ.) for his experimental supports. 
The crystal structures were visualized by VESTA software \cite{vesta}. 
The crystallographic data of \lBEDSe\ at R. T. ($293$~K) and $110$~K can be obtained free of charge from the Cambridge Crystallographic Data Centre \cite{ccdc}. 
This work was in part supported by Hokkaido University, Global Facility Center (GFC), Advanced Physical Property Open Unit (APPOU), funded by MEXT under ``Support Program for Implementation of New Equipment Sharing System'' (JPMXS0420100318). 
This work was also partly supported by the Japan Society for the Promotion of Science KAKENHI Grants No. 20K14401, No. 19K03758, No. 19K14641 and No. 21K03438. 

%
\end{document}